\documentclass[aps,prb,preprint,superscriptaddress,showpacs]{revtex4}

\usepackage{textcomp}
\usepackage{amsmath}
\usepackage{graphicx}
\usepackage{color}

\newcommand{\jmmapprox}{\sim \! \!}
\newcommand{\jmmemph}[1]{\textit{#1}}

\begin{document}

% ******************************
% Title of paper
% ******************************
\title{Nuclear Quantum Effects and Nonlocal Exchange--Correlation Functionals Applied to Liquid Hydrogen at High Pressure}

% ******************************
% Authors / affiliations
% ******************************
\author{Miguel A. Morales}
\email[]{moralessilva2@llnl.gov}
\affiliation{Lawrence Livermore National Laboratory, Livermore, California 94550, USA}

\author{Jeffrey M. McMahon}
%\email[]{mcmahonj@illinois.edu}
% \affiliation{The Institute for Condensed Matter Theory, University of Illinois at Urbana-Champaign, Urbana, Illinois 61801, USA}
\affiliation{Department of Physics, University of Illinois at Urbana--Champaign, Urbana, Illinois 61801, USA}

\author{Carlo Pierleoni}
%\email[]{carlo.pierleoni@roma1.infn.it}
\affiliation{Department of Physical and Chemical Sciences, University of L'Aquila and CNISM UdR L'Aquila, Via Vetoio, I-67010 L'Aquila, Italy}

\author{David M. Ceperley}
%\email[]{ceperley@illinois.edu} 
\affiliation{Department of Physics, University of Illinois at Urbana--Champaign, Urbana, Illinois 61801, USA} 

% ******************************
% Date
% ******************************
\date{\today}

% ******************************
% Abstract
% ******************************
\begin{abstract}
Using first-principles molecular dynamics, we study the influence of nuclear quantum effects (NQEs) and nonlocal exchange--correlation density functionals (DFs) near molecular dissociation in liquid hydrogen. NQEs strongly influence intramolecular properties, such as bond stability, and are thus an essential part of the dissociation process. Moreover, by including DFs that account for either the self-interaction error or dispersion interactions, we find a much better description of molecular dissociation and metallization than previous studies based on classical protons and/or local or semi-local DFs. 
We obtain excellent agreement with experimentally measured optical properties along pre-compressed Hugoniots, and while we still find a first-order liquid--liquid transition at low temperatures, transition pressures are increased by more than $100$ GPa. 
\end{abstract}


%\keywords{}

\maketitle

%\section{Introduction}

Hydrogen, being the most abundant element in the Universe, has a prominent role in planetary science.  
Considerable attention has thus been given to the study of its phase diagram at high pressure, 
both experimentally \cite{Nellis06,Deemyad08,Eremets09,Subramanian11,H_metallization_Eremets-NatMat-2011,Howie12,Zha12} and via first-principles (FP) simulations. The latter have been particularly important, in many cases being instrumental in providing the correct interpretation of conflicting experimental results. This is well exemplified, for instance,
by the experimental controversy over the maximum compression along the principal Hugoniot \cite{Collins98,DaSilva97,Belov02,Boriskov03, Grishechkin04,Knudson04,Knudson09,Hicks09}; simulations overwhelmingly favor one with a maximum compression of $\jmmapprox 4.3$--$4.4$ \cite{Lenosky00,Militzer00,Dejarlais03}, 
in agreement with experiments using magnetic implosions at the Z pinch \cite{Knudson04,Knudson09} and converging explosive-driven shock waves \cite{Belov02,Boriskov03, Grishechkin04}. Another example of the predictive capability of FP simulations is that of a maximum in
the melting line of the molecular solid by FP molecular dynamics (FPMD) \cite{Bonev04},  subsequently 
confirmed by measurements \cite{Deemyad08,Eremets09,Subramanian11}. Unfortunately though, FP methods still employ questionable approximations that could affect their predictions, especially when effects occur near-simultaneously, such as metallization and molecular dissociation.

\begin{figure}[t]
     \includegraphics[scale=0.34]{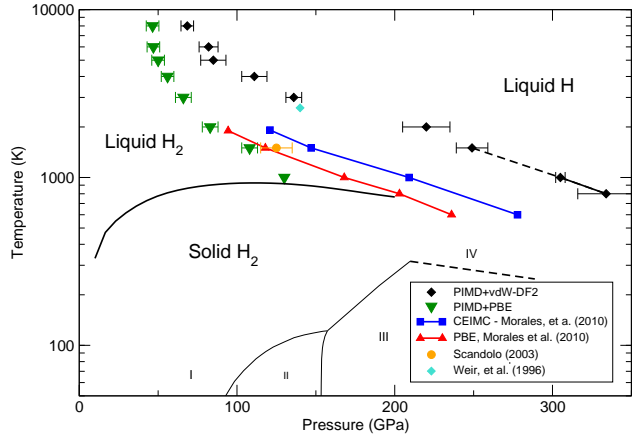}
     \caption{(Color online) Phase diagram of hydrogen. Diamonds (black) and downward-triangles (green) represent PIMD+vdW-DF2 and PIMD+PBE state-points, respectively, at which the electronic conductivity reaches 2000 ($\Omega$ cm)$^{-1}$, which separates the insulating and conducting regimes. Upward-triangles (red) and squares (blue) are results from previous studies using classical protons in FPMD+PBE and coupled electron--ion Monte Carlo simulations \cite{Morales10,Liberatore11}, respectively. The orange circle is the original prediction for the location of the LLPT from Scandolo \cite{Scandolo03} using FPMD within the local density approximation, and the turquoise diamond is the experimental measurement of minimum metallic conductivity by Weir, \emph{et al.} \cite{Weir96} }
     \label{fig:phase-diag}
 \end{figure}

The emerging picture of molecular dissociation in liquid hydrogen, as suggested by FP simulations, is shown in Fig.\ \ref{fig:phase-diag}.
Below a critical temperature of $\jmmapprox1500$--$2000$ K, this occurs through a first-order liquid-liquid phase transition (LLPT) between an insulating molecular liquid and a conducting atomic-like liquid. The LLPT is characterized by a discontinuous change in the electrical conductivity with increasing pressure combined with a small volume discontinuity. Above the critical temperature, the electrical conductivity and dissociation is then continuous with increasing pressure. While different levels of theory agree qualitatively regarding the LLPT \cite{Delaney06,Morales10,Liberatore11,Lorenzen10}, the location of it is still a matter of debate. 

Obviously, the variability in results must arise from approximations employed in the underlying numerical methods. The two main ones typically employed (at least in the case of high-pressure hydrogen) are the neglect of nuclear quantum effects (NQEs) and, in density-functional theory (DFT) studies, the inability to fully treat electronic correlation (e.g., dispersion interactions) as well as approximations to exchange; the latter typically resulting in a strong underestimation of band gaps \cite{stadele00}.

The neglect of NQEs in FP simulations is typically (but not always \cite{Biermann98a,Biermann98b,Kitamura00,Pierleoni04,Pierleoni06,Morales09b,Morales10}) employed, due to increased computational demands.
Neglecting them, however, has an important effect on the calculated LLPT transition pressures. The use of classical protons leads to an overestimation, since the high-frequency vibrations of the molecular bond leads to a large zero-point motion (ZPM), an effect which is much smaller in the atomic phase (at least right at dissociation). 
This effect has been clearly shown by previous path-integral molecular dynamics (PIMD) simulations using the Perdew--Burke--Ernzerhof (PBE) density functional (DF) \cite{Morales10}, where the LLPT is decreased by $\jmmapprox 40$ GPa with respect to FPMD+PBE (i.e., classical protons). However, this would predict a transition pressure around $P_t=130$ GPa at temperature $T_t=1000$ K, a result not supported by current experiments \cite{Deemyad08,Eremets09,Subramanian11}. There are thus additional approximations which are likely to blame.

The vast majority of DFs used in DFT simulations of hydrogen have been based on either the local density approximation or the semi-local generalized gradient approximation. These, however, underestimate the band gap by 1--2 eV \cite{stadele00}. This means that the metallization pressure,  directly related to the dissociation process and thus the location of the LLPT,  will also be underestimated.
As should be clear from this discussion, neglecting NQEs \jmmemph{overestimates} the transition pressure, while using a local or semi-local DF \jmmemph{underestimates} it. Therefore, the two errors partially cancel, providing a reasonable prediction in this particular instance.
Such a cancellation is not always as fortunate, however, as indicated by the calculation of other properties near metallization, such as melting [see the Supplementary Material (SM) \cite{SM}]. Therefore, rigorous simulations including NQEs with local or semi-local DFs must be carried out with caution.

In this Letter, we present results from FP simulations based on PIMD to treat NQEs, but using nonlocal DFs in DFT. These calculations remove one of the most significant approximations made in a number of previous simulations (classical protons), while at the same time improve over another equally important and heretofore less-considered approximation (local or semi-local DFs). Such calculations allow us to study molecular dissociation in hydrogen with previously unattainable accuracy.

%%%%%%%%%%%%%%%%%%%%%%%%%%%%%%%%%%%%%%%%%%%%%%%%%%
%\section{Computational Details}
%%%%%%%%%%%%%%%%%%%%%%%%%%%%%%%%%%%%%%%%%%%%%%%%%%

Simulations were performed via DFT, and we focused on two nonlocal DFs. We first chose to use the Heyd--Scuseria--Ernzerhof (HSE) DF \cite{Heyd03}, which is known to have a very small self-interaction error \cite{Henderson11}. We also performed simulations with the vdW-DF2 DF \cite{Dion04,Thonhauser07,Roman-Perez09,Lee10}, which provides a reasonable description to exchange (for a semi-local functional), but moreover provides an improved description of nonlocal correlation (dispersion interactions) in DFT. 
Simulations with the former were performed with VASP \cite{VASP} and the latter with a modified version of Quantum ESPRESSO (QE) \cite{QE}. 
A time-step of 8 (a.\ u.)$^{-1}$ was used in all simulations, and the path integrals (PIs) were discretized with a Trotter time-step no larger than $0.000125$ K$^{-1}$. After an equilibration periods of $\jmmapprox 0.25$ ps, statistics were gathered for simulation times of $\jmmapprox 1.5$--$2.0$ ps, corresponding to $\jmmapprox 6500$--$9000$ time steps. A Troullier--Martins norm conserving pseudopotential \cite{Troullier91} with a core radius of $r_c = 0.5$ a.\ u.\ was used to replace the bare Coulomb-potential of hydrogen in the QE simulations; a PAW \cite{Kresse99} was used in VASP. System sizes ranged from $128$--$432$ atoms (a large number of atoms has been previously shown to be required for the proper description of the dissociation transition in liquid hydrogen at lower temperatures \cite{Morales10}). All simulations were performed at the Gamma point. The simulations with QE were performed with a plane-wave cutoff of 1224 eV, while the simulations with VASP were performed with a plane-wave cutoff of 250 eV and ``Accurate" settings.  Finite-temperature effects on the electrons were taken into account by using Fermi--Dirac smearing \cite{VASP}.  
While most PIMD simulations were performed with the standard primitive approximation \cite{Ceperley95}, the simulations of 432 atoms at temperatures of $1000$ and $1500$ K utilized the accelerated PIMD method of Ceriotti, \emph{et al.} \cite{Ceriotti}, based on a generalized Langevin dynamics (GLE) and the Born--Oppenheimer approximation. The use of the PI+GLE method was carefully tested under the relevant pressure and temperature conditions, in order to guarantee proper convergence \cite{McMahon12b}.

%%%%%%%%%%%%%%%%%%%%%%%%%%%%%%%%%%%%%%%%%%%%%%%%%%
%\section{Results}
%%%%%%%%%%%%%%%%%%%%%%%%%%%%%%%%%%%%%%%%%%%%%%%%%%
 
As detailed further in the SM \cite{SM}, we first compare the structure and equation of state (EOS) data for systems of classical protons calculated using vdW-DF2 and HSE (the latter more computational demanding) \footnote{Notice that HSE simulations are very sensitive to the choice of k-points and the size of the simulation cell, which means that the transition pressures will depend on the particular details of the simulation. This is not the case for the other DFs employed in this work. While the precise location of the transition using HSE is a matter of future work, away from the transition the dependence on simulation details should be considerably smaller. }.
The two DFs provide pair correlation functions (PCFs) in very good agreement, even though pressures from HSE are $\jmmapprox 7\%$ smaller. This means that the LLPT lines predicted by the two methods will only be \jmmemph{slightly} different. This should be compared to the predictions of PBE, which results in a very large disagreement relative to either nonlocal DF.
 
\begin{figure}[t]
     \includegraphics[scale=0.65]{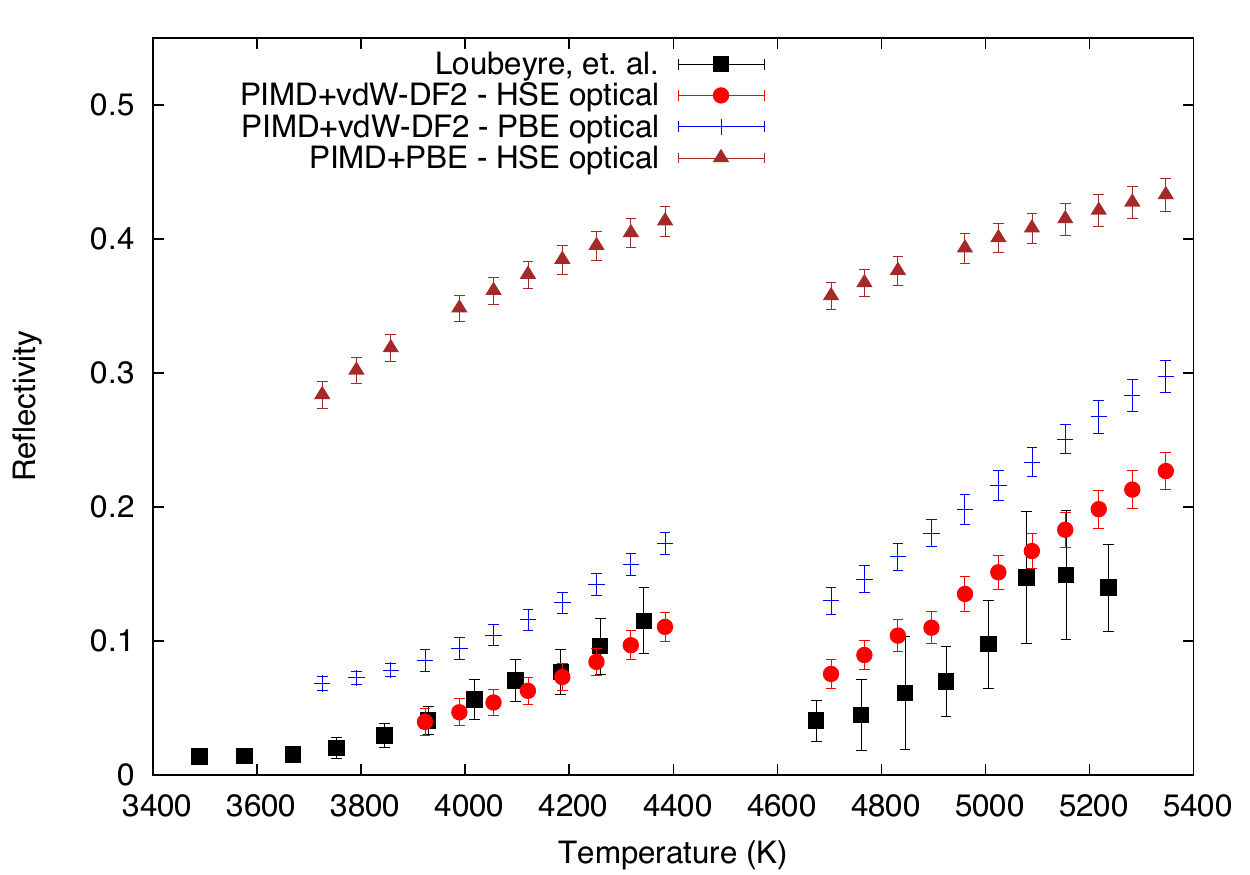}
     \caption{(Color online) Comparison of the measured reflectance along pre-compressed Hugoniots from Loubeyre \emph{et al.}, \cite{Loubeyre04} to those calculated in this work. The reflectivity was calculated on the pressure-temperature curves reported by Loubeyre \emph{et al.} The two sets of data correspond to experiments with different initial conditions. Red circles and blue crosses correspond to PIMD simulations with vdW-DF2 and with the optical properties being calculated using either HSE or PBE, respectively. The influence of the underestimation of the bandgap on optical properties calculated with PBE is clear. Green stars correspond to PIMD simulations with PBE and optical properties calculated with HSE. In this latter case, the influence on the optical properties caused by structural differences (from the trajectories) is far more important.}
     \label{fig:reflectivity}
 \end{figure}

We performed an extended set of FPMD simulations with classical protons as well as PIMD simulations using vdW-DF2 at temperatures ranging from $800$ to $8000$ K at densities in the region near molecular dissociation.  
At every density and temperature, optical properties were calculated within the Kubo--Greenwood formulation \cite{KG}, by performing excited state calculations on $15$ statistically-independent proton configurations. Note that trajectories and optical properties were not necessarily calculated using the same DF, the former which is denoted in the following notation. Results are reported in Fig.\ \ref{fig:reflectivity}, in comparison to experimental results for hydrogen along pre-compressed Hugoniots \cite{Loubeyre04}. We first note that the reflectivity data relative to configurations obtained with PIMD-vdW-DF2 are in good agreement. The quality of the prediction is affected by the DF used in the optical calculation, HSE providing an excellent agreement with experiments, having a slightly lower reflectivity than PBE. As discussed above though, this is not unexpected, due to the well known bandgap problem of local and semi-local DFs.
On the other hand, reflectivity results from configurations obtained with PIMD-PBE are $\jmmapprox 3$ times larger than the experimental values, even when the optical calculations are performed with HSE. This effect likely derives from the strong tendency of PBE to favor delocalized electronic states combined with its poor treatment of dispersion interactions, which probably results in inaccurate proton statistical configurations, and thus the metallization and LLPT process altogether. 

It is important to mention that the above simulation data agrees very well with the SESAME EOS \cite{Kerley03,Loubeyre04}, the latter used to convert experimental shock velocity data to pressure, density, and temperature.
 \begin{figure}[t]
     \includegraphics[scale=0.70]{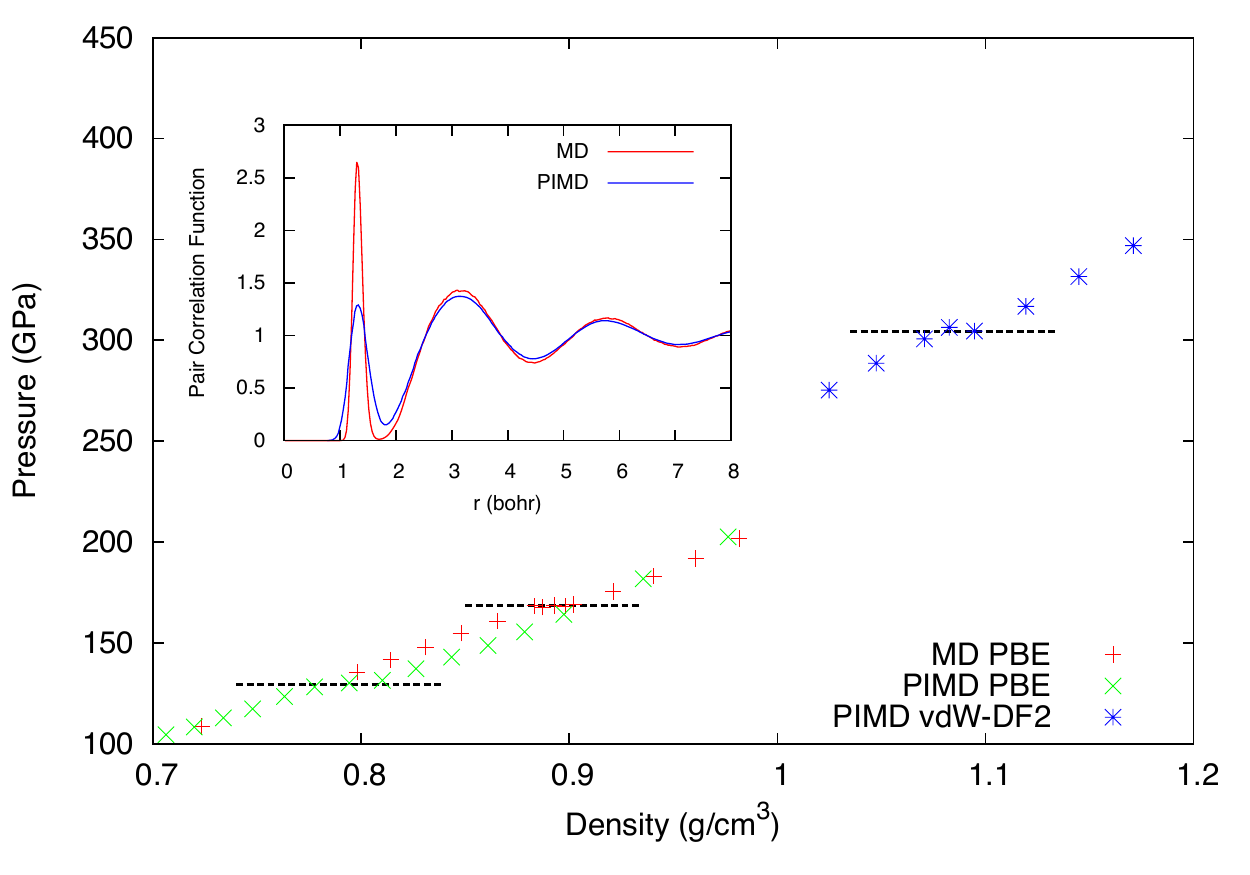}
     \caption{(Color online) Comparison of pressure isotherms at 1000 K for classical protons with PBE (red), 
     quantum protons with PBE (green), and quantum ions with vdW-DF2 (blue). The horizontal dashed black lines show the pressure plateau 
     where the LLPT takes place. Inset: Comparison of PCFs between simulations of classical (red) 
     and quantum (blue) protons using vdW-DF2 at a density of $\rho \sim 0.88 \ $g/cm$^3$.  }
     \label{fig:pvsrho}
 \end{figure}
For example, our present thermodynamic data (PIMD-vdW-DF2) predicts a pressure only slightly higher ($\jmmapprox3$--$5\%$) than SESAME in the relevant density range. 
Further, along the $T = 5000$ K isotherm, the agreement is better than 1\% for pressures in the range of the experiments ($30$--$60$ GPa).

Figure \ref{fig:pvsrho} shows a comparison of pressure versus density along the $T=1000$ K isotherm for both FPMD and PIMD simulations using either PBE \cite{Morales10} or vdW-DF2. Notice that both DFs show a plateau in the pressure, a clear indication of a first-order LLPT. 
There is, however, a further qualitative similarity in that the transition occurs between an insulating molecular liquid and 
a conductive atomic-like liquid. There is a large quantitative difference in the transition pressures. 
The inset of Fig.\ \ref{fig:pvsrho} shows a comparison of the PCF between FPMD and PIMD simulations using vdW-DF2. As can be seen, NQEs have a strong influence on the
properties of the molecular peak, ZPM producing a wider distribution of bond distances. This results in a destabilization of the molecular state,  
explaining the lower transition pressures. (Notice that the primary vdW-DF2 results shown in the figure are performed with PIMD, 
so systems of classical protons are expected to exhibit even higher transition pressures, above 365 GPa).

\begin{figure}[t]
     \includegraphics[scale=0.3]{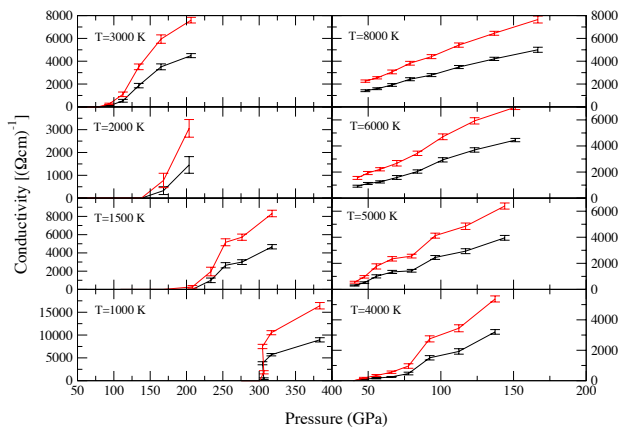}
     \caption{(Color online) Electronic conductivity as a function of pressure, along various isotherms. Results calculated with both HSE (black-lower) and PBE (red-higher) are shown for comparison. }
     \label{fig:sigma}
 \end{figure}

Figure \ref{fig:sigma} shows the electronic conductivity as a function of pressure along various isotherms, comparing both PBE and HSE. Note that in both cases, proton configuration were generated with vdW-DF2.
Notice also that while the conductivity values differ between HSE and PBE, they nonetheless agree on the \jmmemph{existence} of a jump at $T=1000$ K.  
 
Returning to Fig.\ \ref{fig:phase-diag}, a schematic phase diagram of hydrogen in the regime of molecular 
dissociation and below $T=6000$ K can be seen. The previously
reported LLPT, obtained with classical protons and either from FPMD+PBE
or coupled electron--ion Monte Carlo (CEIMC) \cite{Morales10} calculations are shown \footnote{The CEIMC method is an alternative FP simulation method based on a quantum Monte Carlo treatment of the electronic potential energy surface, combined with a classical or quantum treatment of the ions using traditional Monte Carlo methods.}.
Both vdW-DF2 (present work) and CEIMC calculations show a considerable increase in the transition pressures with respect to PBE, with those from vdW-DF2 being considerably higher. 
Above the critical point, state-points of an electronic conductivity of $\sigma=2000 ~ (\Omega \text{cm})^{-1}$, separating the insulating from metallic liquid \cite{Weir96}, are also reported using either vdW-DF2 or PBE. 
Loubeyre \textit{et al.} \cite{Loubeyre04} reported that the metal-to-insulating threshold was located at conditions of 10\% reflectivity, since according to the Drude model, this corresponds to an ionization of 1\%. The present criterion for metallic behavior is different though. For example, from our reflectivity data, a minimum metallic conductivity of $\sigma=2000 ~ (\Omega \text{cm})^{-1}$ corresponds to a reflectivity of $\jmmapprox 0.35$--$0.40$ which is closer to $70\%$ of its saturation value ($\jmmapprox 0.6$). This explains why our threshold line is in apparent disagreement with the experimental points reported in Ref.\ \onlinecite{Loubeyre04}. In fact, at conditions of 10\% reflectivity, close to the pre-compressed Hugoniot, we observe conductivities on the order of $\sigma=100$--$500 ~ (\Omega \text{cm})^{-1}$. 
Figure \ref{fig:phase-diag} also shows the result from the reverberation shock compression of S. Weir, \emph{et al.} \cite{Weir96}.
While the temperature was not measured therein experimentally, but rather estimated using a
model EOS, and the error bars were rather large, it is nonetheless clear that the presented results of the location of the LLPT and the dissociation regime at higher temperatures agree rather well.

%\section{Discussion}

While almost all FP simulation methods agree qualitatively on the existence of a first-order LLPT in high-pressure hydrogen \cite{Morales10,Lorenzen10}, its precise location depends on the approximations employed. The results reported above clearly show that NQEs and non-local DFs in DFT play an important role in the description of molecular dissociation and metallization. The two DFs considered (HSE and vdW-DF2) were originally developed with the goal of addressing significant limitations of local and semi-local DFs in DFT. 
HSE, on the one hand, was developed to reduce self-interaction errors in PBE in its applications to solids \cite{Heyd03}. Such errors lead to a strong tendency to favor delocalized electronic states, which in turn lead to an underestimation of bandgaps by as much as $1$--$2$ eV (in hydrogen) \cite{stadele00}. This leads to a serious underestimation of the metallization pressures in both liquid and solid phases, and a tendency to favor metallic states (e.g., solid structures). vdW-DF and its improved version vdW-DF2 (employed in this work), on the other hand, were developed to account for nonlocal electron correlations, such as dispersion interactions in DFT. The presented results indicate that, at least close to dissociation, both HSE and vdW-DF2 DFs produce very similar structures in liquid hydrogen. Since the physical effects addressed by both DFs are not directly related to each other, and that both effects are expected to be relevant in the molecular phase, it is important to recognize that the LLPT pressures might still change if a DF which combines both hybrid exchange and non-local correlation were to be employed.

The goal of this Letter was not to predict which functional (HSE and vdW-DF2, etc.) is more accurate, since answering that question requires the use of more accurate methods\footnote{Preliminary results in solid molecular hydrogen show that PIMD+vdW-DF2 is able to reproduce the measured bandgap of the solid, while PIMD+HSE predicts gaps that are $\jmmapprox 1$ eV too low \cite{Morales12}. These results provide evidence in favor of the higher transition pressures predicted by the PIMD+vdW-DF2 simulations.}. We can however mention several possibilities that explain the observed behavior, the reasonable agreement between either nonlocal DF as well as their large disagreements with PBE. Both DFs predict shorter molecular bonds compared to PBE; in the limit of low density, the bond length predicted by vdW-DF2 agrees very well with measured values while that of PBE is overestimated by $\jmmapprox 3\%$ \cite{Morales12}. This is obviously an important factor on dissociation. Second, the exchange portion of the vdW-DF2 functional was constructed to reproduce exact-exchange results \cite{Lee10}, which may explain its similarity to HSE. Finally, both dispersion interactions and a reduced self-interaction will lead to a more stable molecular state. An even more promising alternative to DFT is the use of quantum Monte Carlo first-principles methods, for example CEIMC \cite{Morales10}, using accurate trial wave functions, such as those constructed from HSE orbitals. We must also recognize that, while the use of the vdW-DF2 DF made large improvements in the description of molecular dissociation in hydrogen near the LLPT, standard semi-local DFs like PBE have been shown to be successful in describing other materials when combined with the HSE DF for the calculation of optical properties \cite{Knudson12}.

While the presented results are obviously important to high-pressure hydrogen, they also suggest that NQEs will have a strong influence on the bonding properties of other hydrogen-rich materials, particularly in the description of transitions between phases with different bonding characteristics. Although this has been demonstrated in some cases, such as high-pressure ice \cite{Marx98}, and similar effects could be present in many other materials (e.g., methane and ammonia), the simultaneous and proper inclusion of NQEs has been largely ignored in the field of FP simulations.  With the development of efficient PI methods, such as the PI+GLE \cite{Ceriotti}, and faster computers, we expect that future simulations will routinely include them, neither them neglecting or employing approximations at the harmonic level. 
 
%\section{Conclusion}

In conclusion, we have studied liquid hydrogen at high pressure using nonlocal exchange-correlation DFs in DFT, namely HSE and vdW-DF2. Both produce similar descriptions of the liquid, with large increases in the LLPT pressures (where molecular dissociation occurs) by more than $100$ GPa (at lower temperatures), from earlier studies using PBE. We also presented a detailed study of the influence of NQEs in the LLPT in combination with vdW-DF2. Remarkable agreement with experiment was observed for optical properties along pre-compressed Hugoniots, as well as with reverberating shock compressed measurements at low temperatures.  The improved description further confirms the existence of a first-order LLPT between an insulating molecular liquid and a conductive atomic-like state at high pressures and below a critical temperature of $T_c \approx 1500$--$1000$ K. Since this work presents a highly-accurate prediction of the location of the LLPT in hydrogen, it should serve as a clear goal for future experimental and theoretical works in this field.

% ******************************
% ACKNOWLEDGMENTS
% ******************************
\begin{acknowledgments}
The authors would like to thank Sebastien Hamel, Alfredo Correa and Eric Schwegler for insightful discussions. 
M.\ A.\ M.\ was supported by the U.S. Department of Energy at the Lawrence Livermore National Laboratory under Contract DE-AC52-07NA27344 and by LDRD Grant No. 10-ERD-058. J.\ M.\ M.\ and D.\ M.\ C.\ were supported by DOE DE-FC02-06ER25794 and DE-FG52-09NA29456. C.\ P.\ was supported by the Italian Institute of Technology (IIT) under the SEED project Grant 259 SIMBEDD. 
Computer time was provided by the US DOE-INCITE program, Lawrence Livermore National Laboratory through the 6th Institutional Unclassied Computing Grand Challenge program and by EU-PRACE project number 2011050781. This research was also supported in part by the National Science Foundation through XSEDE resources provided by NICS under grant number TG-MCA93S030. 
\end{acknowledgments}

\end{document}